\documentclass[sigconf]{acmart}

\AtBeginDocument{%
  \providecommand\BibTeX{{%
    \normalfont B\kern-0.5em{\scshape i\kern-0.25em b}\kern-0.8em\TeX}}}

\usepackage{amsmath}
\usepackage{xspace}
\usepackage{balance}
\usepackage{enumitem}
\usepackage{ifthen}
\usepackage[skip=2pt]{caption}
\usepackage{subcaption}
\usepackage{hyperref}
\usepackage{multirow}
\usepackage{graphicx}
\usepackage{framed}
\usepackage{microtype}
\usepackage{xcolor}
\usepackage{balance}
\usepackage{multicol}
\usepackage{listings}
\usepackage{tabularx}
\usepackage{ragged2e}
\usepackage{array}
\usepackage{makecell}
\usepackage{multirow}
\usepackage{tcolorbox}
\usepackage{paralist}

\lstdefinestyle{mystyle}{
  backgroundcolor=\color{white},commentstyle=\color{green},
  keywordstyle=\color{purple},
  numberstyle=\tiny\color{gray},
  stringstyle=\color{purple},
  basicstyle= \small,
  breakatwhitespace=false,
  breaklines=true,
  captionpos=b,
  keepspaces=true,
  numbers=left,
  numbersep=5pt,
  showspaces=false,
  showstringspaces=false,
  showtabs=false,
  tabsize=2,
}
{\endMakeFramed}

\newtcolorbox{resultbox}{colback=lightgray, arc=0.5mm, top=1mm, bottom=1mm, left=1mm, right=1mm}

\newcommand{\ie}{\textit{i.e.,}\xspace}
\newcommand{\eg}{\textit{e.g.,}\xspace}

\newboolean{showcomments}
\setboolean{showcomments}{true}

\ifthenelse{\boolean{showcomments}}
{\newcommand{\nb}[2]{
		\fbox{\bfseries\sffamily\scriptsize#1}
		{\sf\small$\blacktriangleright$\textit{#2}$\blacktriangleleft$}
	}
}
{\newcommand{\nb}[2]{}
}

\acmSubmissionID{}

\copyrightyear{2026}
\acmYear{2026}
\setcopyright{cc}
\setcctype{by}
\acmConference[ICSE-FoSE]{2026 IEEE/ACM 48th International Conference on Software Engineering: Future of Software Engineering }{April 12--18, 2026}{Rio de Janeiro, Brazil}
\acmBooktitle{2026 IEEE/ACM 48th International Conference on Software Engineering: Future of Software Engineering (ICSE-FoSE), April 12--18, 2026, Rio de Janeiro, Brazil}
\acmPrice{}
\acmDOI{10.1145/3793657.3793881}
\acmISBN{979-8-4007-2479-4/2026/04}
\begin{document}

\title{Future of Software Engineering Research: \\ The SIGSOFT Perspective}

\author{Massimiliano Di Penta}
\affiliation{%
 \institution{University of Sannio}\city{Benevento}\country{Italy}
}
\email{dipenta@unisannio.it}

\author{Kelly Blincoe}
\affiliation{%
 \institution{University of Auckland}\city{Auckland}\country{New Zealand}
}
\email{k.blincoe@auckland.ac.nz}

\author{Marsha Chechik}
\affiliation{%
 \institution{University of Toronto}\city{Toronto, ON}\country{Canada}
}
\email{chechik@cs.toronto.edu}

\author{Claire Le Goues}
\affiliation{%
 \institution{Carnegie Mellon University}\city{Pittsburgh, PA}\country{United States}
}
\email{clegoues@cs.cmu.edu}

\author{David Lo}
\affiliation{%
 \institution{Singapore Management University}\city{Singapore}\country{Singapore}
}
\email{davidlo@smu.edu.sg}

\author{Emerson Murphy-Hill}
\affiliation{%
 \institution{Microsoft, Inc.}\city{Sunnyvale, CA}\country{United States}
}
\email{emerson.rex@microsoft.com}

\author{Thomas Zimmermann}
\affiliation{%
 \institution{University of California, Irvine}\city{Irvine, CA}\country{United States}
}
\email{tzimmer@uci.edu}

\renewcommand{\shortauthors}{Di Penta, Blincoe, Chechik, Le Goues, Lo, Murphy-Hill, and Zimmermann}

\begin{abstract} 
As software engineering conferences grow in size, rising costs and outdated formats are creating barriers to participation for many researchers. These barriers threaten the inclusivity and global diversity that have contributed to the success of the SE community. Based on survey data, we identify concrete actions the ACM Special Interest Group on Software Engineering (SIGSOFT) can take to address these challenges, including improving transparency around conference funding, experimenting with hybrid poster presentations, and expanding outreach to underrepresented regions. By implementing these changes, SIGSOFT can help ensure the software engineering community remains accessible and welcoming.
\end{abstract}



\maketitle

\section{Introduction}
\label{sec:intro}
Almost 60 years after the 1968 NATO conference popularized the term ``software engineering", the software engineering (SE) research community is growing, as can be observed by the increasing number of submissions the broad SE conferences received over the past five years.
Additionally, SE research is evolving as it adapts to changing technology and societal needs. This has manifested in, for example, growing attention to topics related to Artificial Intelligence (AI) applications in SE, as well as to SE methods for AI, and also to software security.

As the community grows and evolves, so too must its structures and practices. A healthy research community should complement this natural evolution with an effective, efficient, and sustainable adaptation of its governance mechanisms, and should be able to keep a healthy and respectful environment.

This paper presents the position of the ACM Special Interest Group on Software Engineering (SIGSOFT). Created 50 years ago, SIGSOFT now (co-)sponsors over 20 SE conferences and cooperates with a dozen others. SIGSOFT organizes several activities within the SE community, including:
\begin{compactitem}
\item Managing different types of awards, including the Outstanding Research, Distinguished Service, Influential Educator, Impact Paper, Outstanding Doctoral Dissertation, and Early Career Researcher Awards;
\item Organizing CARES, a committee whose role is to listen to and help people who experience or witness discrimination, harassment, or other ethical policy violations;
\item Organizing webinars with leading SE researchers from industry or academia;
\item Running CAPS, which provides travel and childcare support to conference participants;
\item Providing financial support for the organization of summer/winter schools for graduate students and junior researchers;
\item Providing financial support for student participation through the Mentorship workshop (SMeW) at ICSE;
\item Providing financial support for participation of researchers from diverse geographical areas, for example, through the SIGSOFT Africa initiative, but also through various local SIG sub-committees;
\item Communicating with the community via Software Engineering Notes (SEN), SIGSOFT blog, and social media.
\end{compactitem}

We begin by reviewing the dataset collected by the ICSE 2026 Future of Software Engineering Workshop pre-survey \cite{storey_2025_18217799} to identify the strengths and weaknesses of the SE community. Then we identify areas of improvement in which SIGSOFT could play an effective role.
We looked at the answers to the following questions:
\begin{compactitem}
\item What aspect or aspects of the SE research community work well, and why?
\item What aspect or aspects of the SE research community do not work well, and why?
\item If you could make one change, what would you change, and what outcome from that change would you like to see in the SE research community?
\end{compactitem}

We focused on responses mentioning positive and negative aspects, as well as suggestions for improvement, related to (i) facilitating onboarding and participation, (ii) creating a welcoming and friendly environment, and (iii) transparency and inclusiveness in the community governance. We excluded responses related to the conference's content and topics, as well as the reviewing process. The response coding was conducted through a thematic analysis performed through online spreadsheets. In the following sections, we report and discuss the gathered findings, relating them to ongoing and future SIGSOFT initiatives.

\section{What works well?}
We identified 47 answers related to the investigated topics. 

Firstly, the survey participants appreciated the high level of openness within the community (11 responses). For example, I22 mentioned how \emph{``There's a strong drive to rethink and modernize the discipline. New initiatives, collaborations, and paper formats are emerging that aim to make research more transparent, reproducible, and impactful.''}
I188 is pointed out that
\emph{``The software engineering research community is very welcoming to researchers from other communities \dots"} 
Other respondents indicated that they perceive the community as more open to discussing new research topics than other research communities to which they belong.
On a similar note, participants found SE conferences to be a very welcoming environment (10). 
Also, participants found SE events to be greatly respectful of people, also facilitating diversity 
in all its aspects (8). 

Participants (4) appreciated the geographical distribution, fixed
schedule, and, in general, the dialogue between the three general SE conferences (ICSE, FSE, and ASE), made possible through the SE3 (the coordinating committee of the three conferences) community effort. For example, I90 mentioned \emph{``I like having three conferences in different parts of the world each year.''}
Participants also liked the fact that the community is very willing to improve (3 answers). 

While researchers appreciated that conferences continued to happen (in virtual mode) during the pandemic, 
some respondents (3) highlighted that they greatly value in-person participation, which they found hard to replace with virtual/hybrid alternatives. I63 said \emph{``\dots there are many ways to be sustainable, and this does not necessarily need to include having a hybrid format for conferences.''}

Other appreciated elements were the presence of workshops and small gathering events; the attention to sustainability; the evolving reviewing models; the active attempt to involve volunteers in the conferences' organization/governance processes, a growing community, a better gender diversity than other CS conferences, and the presence and value of community meetings, including the SIGSOFT/TCSE Townhall.

\begin{resultbox}
Respondents appreciate the fact that the SE community is friendly, open, welcoming, and respectful. Conferences are viewed as safe spaces that foster diversity and encourage interdisciplinary collaboration. The geographical distribution of major venues and the community's willingness to evolve are also valued.
\end{resultbox}

\section{What Are the Problems?} 
\label{sec:problems}

Respondents also identified problems they perceive within the SE community. We identified 42 responses relevant to the topics examined in this paper.

Perhaps unsurprisingly, the most frequently highlighted problem is the high cost of conference attendance (9 responses). One participant (I45) noted that during the pandemic, more researchers from developing countries attended SE conferences; however, this trend has since ceased. Others echoed this, indicating how costs penalize participation and inclusiveness.
More generally, some respondents (5) noted that, despite the conference's geographic distribution, certain countries and populations are underrepresented. 
I78 mentioned \emph{``We're still mostly white or Asian, mostly representing the northern and western parts of the world''}.

As conferences continue to grow in size, a conventional format with primarily paper-based presentation sessions may no longer be effective. Therefore, 8 respondents suggested rethinking conference formats. For example, 
 I244 mentioned \emph{"conferences have very short talks, making the work seem very superficial."}. This causes limited interaction and networking, as I19 pointed out \emph{``\dots little opportunity to discuss and share research progress or ideas asynchronously beyond the researchers you interact with already.''}. 


Participants also observed a polarization towards the largest general conferences, especially ICSE, FSE, and ASE (3). I59 brought out the polarization point which has recently penalized specialized venues: ``Big conferences are getting bigger, but small and medium-sized conferences are getting smaller.'' I73 mentioned that this may be due to existing ranking systems, which ``lead to an artificial separation of the `top venues' and the `rest'.''

Three respondents asked for more inclusiveness and transparency in governing bodies, including societies and conference steering/organizing committees. Three respondents also pointed out the constraints imposed by societies (IEEE and ACM) on organizing conferences, considering the society overhead a sort of ``profit'' rather than money that is reinvested in the research community. I207 mentioned \emph{``It is heartbreaking to see how tons of public money goes to publishers who charge for both open access and for conference participation (by means of very high overheads on the expenses).''}

Two participants each identified the following problems: 
\begin{compactitem}
\item The need to find a better form of recognizing reviewers, \eg, through reduced registration fees;
\item Concerns about the introduction of ACM Open Access and the consequent costs for researchers whose organizations do not participate in the ACM Open.
\item The tendency of remote participation and proxy presenters to disrupt the conference experience.
\item The clustering of deadlines in the period from mid-March to early September, notwithstanding the geographic/temporal distribution of conferences. 
\end{compactitem}

Other concerns include the proliferation of conferences, with all conferences becoming too similar; the fact that Asia should no longer be treated as the ``rest of the world" when planning conferences, given its significant contribution in terms of submissions/authors; and the need to improve inter-generational interactions.
\begin{resultbox}
The high cost of conference attendance is the most frequently cited problem, limiting participation and inclusiveness. Respondents also call for rethinking conference formats to enable deeper interaction, greater transparency in society governance and fund allocation, and addressing the underrepresentation of certain regions despite the geographical distribution of venues.
\end{resultbox}

\section{Suggested Improvements} 
\label{sec:improvements}

The most common theme addressed suggested changes to the current publication model (37). There was no clear consensus on what the changes should be, but many expressed frustration with the current model. One of the most common suggestions was to adopt a model where conferences centered on community building and discussions of ongoing work or recent journal articles, rather than published conference proceedings. P181 said \textit{"Change the whole publication process. No more conference publications. Move all these papers to journals."} Other suggestions included merging conferences, increasing deadline frequency, reducing the number of allowed submissions, both reducing and increasing page limits, both reducing the number of small workshops (silos) and breaking up large conferences (to be more focused and interactive), and breaking away from current publishers.

Another common theme was making our community and events more inclusive (21). Many of the comments related to inclusion centered around making our events more accessible through online participation, funding, and diverse conference locations. There were also suggestions for having more open calls for volunteers for organizing and program committees, as well as lower barriers to entry for these roles. In addition, several comments on inclusion suggested that the community needed to be more open to new ideas, different research methods, and unfamiliar researchers. For example, P28 said \textit{"...in my country, there is still a strong fear of trying innovative ideas that might be perceived as challenging the work of well-established and renowned researchers in the field."}


There were some suggestions (4) around increased outreach to improve the visibility of the research conducted by our community. The suggestions included podcasts, recorded talks, social media, and other avenues for sharing research results beyond the academic community for broader impact.  Similarly, there were suggestions for improved communication mechanisms (3) within the research community, including cross-venue informational portals that provide information on conference deadlines, maintain lists of active researchers and their current research topics, and provide additional community communication channels. 

Some suggestions focused on environmental sustainability (3). Suggestions included not requiring conference attendance for publication, holding multi-location conferences to reduce travel-related carbon emissions, and serving vegan food at conferences.  

Last, but not least, there were suggestions to make societal decisions more transparent (1) and to ensure follow-up from surveys like this one (1). 


\begin{resultbox}
Respondents most frequently suggested rethinking the publication model, from journal-first approaches to restructuring conferences around community building. Improving inclusivity---through online participation, diverse locations, and lower barriers to organizing roles---was another major theme, alongside open science, outreach, and environmental sustainability.
\end{resultbox}

\section{The SIGSOFT Perspective and Role, and what can be done better}
\label{sec:sigsoft}


\textbf{Actions that are having success.} 
Survey respondents unambiguously appreciated the welcoming, respectful, and friendly environment at SE conferences. To ensure this continues, SIGSOFT is strongly committed to supporting initiatives such as CARES and the anti-harassment policies.

\textbf{Organizing a SIGSOFT conference is costly and has a lot of constraints.}
SIGSOFT would like to improve its transparency regarding the allocation of conference funds.
For each \$100 a conference spends, $\sim$\$14 goes to ACM to cover its personnel and office expenses and to provide services to the conferences, such as insurance, submission systems (HotCRP, EasyChair), and other tools such as plagiarism checks (iThenticate). About \$6 and the surplus (if any) goes to SIGSOFT as  
funding necessary to perform all the initiatives outlined in the introduction, and which contributed (as mentioned above) to the positive elements of the SE conferences, including (i) travel support to students through CAPS and to participants of programs such as the African one, (ii) prizes and travel support to recipients of SIGSOFT awards, (iii) conference infrastructure (Researchr license) and (iv) reinvestment of surplus in future editions of conferences or coverage of conference losses when they occur -- which, sadly, is frequent in the past several years (otherwise, the organizers would have to take on the financial risk themselves).

While ACM budgets are publicly available, SIGSOFT will aim to make its own budget even more transparent by publishing it on the SIGSOFT portal. 

Organizing a conference with a society entails rigorous administrative activities, including the preparation and prior approval of a detailed budget that must meet several constraints. While this appears like a waste of time, it helps ensure that (i) the conference maintains sustainable costs and registration fees, and (ii) the risk of loss is limited. The latter is essential to allow for reinvesting money in future conferences and activities, as mentioned above.

\textbf{Favoring in-person conference participation.}
In general, the remote participation in conferences during the pandemic has not been a success, and most participants are happy to be back networking in person with their peers. 
However, there are many other reasons, still prevalent today, that prevent a portion of the community from participating in conferences, including increasing costs (combined with low funding), geopolitical reasons, or a variety of family or personal reasons. SIGSOFT is actively working to improve the hybrid/remote component of conferences. 

\begin{figure}
    \centering
    \includegraphics[width=0.8\linewidth]{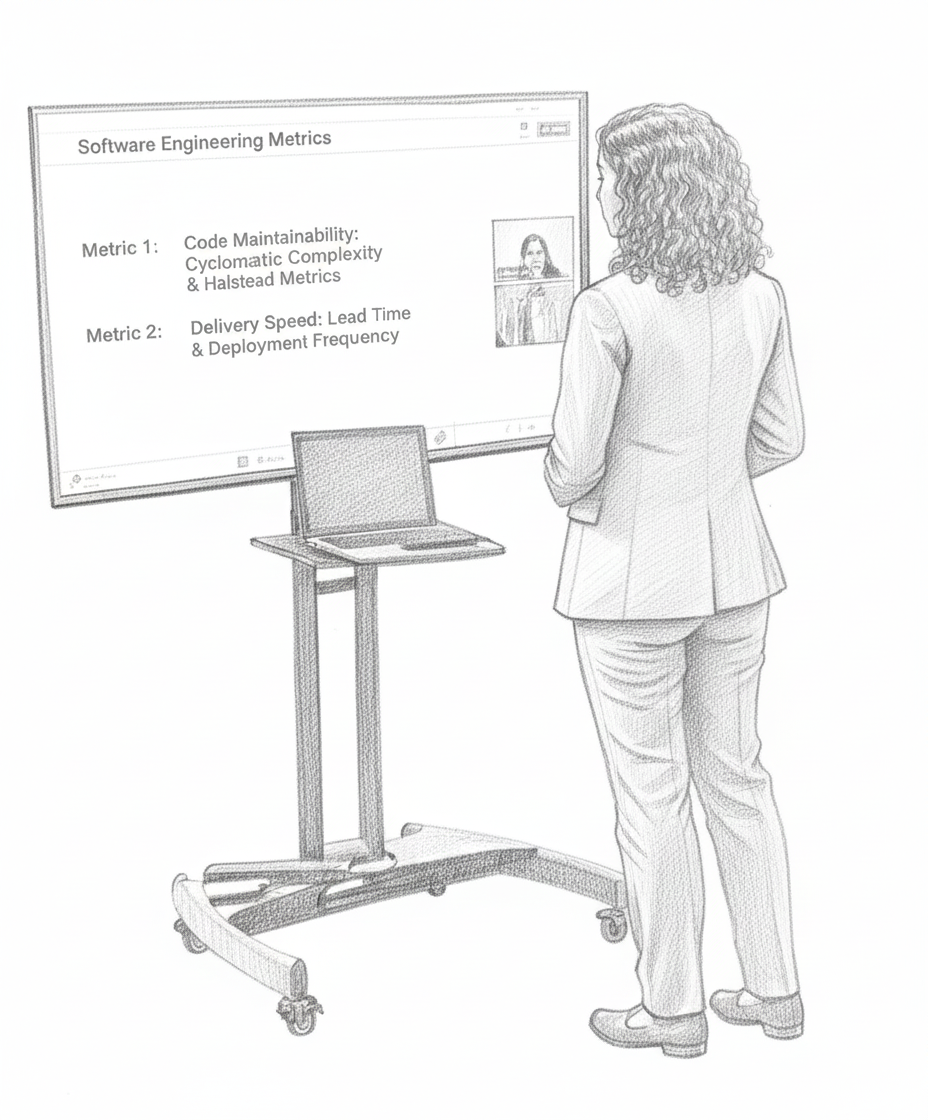}
    \caption{A hybrid poster presentation (credit: Gemini).}
    \label{fig:poster}
    \vspace{-3mm}
\end{figure}

One promising direction is hybrid poster presentations,\footnote{For a detailed proposal on hybrid poster presentations, see \url{https://docs.google.com/document/d/1nA8xduIiic-dUtA7sQbEI3863jMcbzyrVZEOVCvZOAk/edit}.} where remote presenters can engage with in-person attendees through dedicated stations at the venue~\cite{moss2025accessible}. Initial experiments at VL/HCC 2025 tested two approaches: (1) a TV-based video call station using Microsoft Teams, where the remote presenter appears on a large screen and shares their poster (Figure~\ref{fig:poster}), and (2) a VR headset station using Meta Horizons Workrooms, where attendees wear a headset to interact with the presenter's avatar in a virtual environment. Both approaches showed promise: the TV setup offered familiarity and good audio/video quality, while the VR setup provided better audio isolation in noisy environments and a more immersive experience. However, challenges remain, including the need for active facilitation to encourage attendee engagement, technical setup complexity for VR, and ensuring natural interaction. Future experiments will explore scaling these approaches, improving camera placement and audio quality, and framing remote presentations as premium opportunities rather than second-tier alternatives.

Another dimension on which SIGSOFT would like to increase its actions (money permitting) is outreach to countries from the Global South. This implies keeping the SIGSOFT Africa initiative active (to increase participation from the least represented continent) and attempting to replicate similar programs in other parts of the world.

\textbf{On the evolving ecosystem of software engineering conferences, and journals.}
The SE conference ecosystem is evolving: larger, broader conferences are expanding in size, while some smaller ones are contracting.
There are different models that the SE community can consider in the future to accommodate such an evolution, while also maintaining the identity of the various sub-communities. This may include:
\begin{compactitem}
\item Encouraging collocation, to facilitate the organization, limit travel costs, and allow different sub-communities to meet;
\item Separate reviewing process from conference organization, and have papers accepted by a unified reviewing committee to be presented in one of the next conferences; or
\item Move entirely to a journal-first model, as some other communities have done. 
\end{compactitem}

\textbf{The role of a SE community in the era of generative AI}
Anybody who has attended SE conferences in the last five years has witnessed a clear and increasing prevalence of papers related to "Artificial Intelligence (AI) with and for SE". We would like to discuss separately the promises and perils related to AI for SE and SE for AI articles.

As for \textbf{AI for SE,} we are observing how nowadays AI-intensive techniques and, in particular, foundation models allow for either providing better solutions to problems the SE community already modeled and solved in the past, or to solve problems for which, so far, automation was not possible or not even conceivable. 
While this is great, it also highlights crucial questions: What is the specific role and contribution of SE researchers and experts in this context? Would AI researchers be able to "do it better", given their deeper knowledge of the underlying technology, or would SE researchers be better equipped, given their domain knowledge? How to keep the best work and researchers working in the intersection of SE and AI to remain in or to move to the SE community, creating impact on SE practice that is more and more automated by AI?

As for \textbf{SE for AI}, \ie the development of SE techniques specifically tailored for AI models and AI-intensive systems, the promises and perils are even more challenging.  On the one hand, the peculiar characteristics of AI-intensive systems are creating a vast array of opportunities for SE, considering the work conducted so far in designing, testing, or repairing those systems. At the same time, the boundary between some pieces of work developed in this area and work typically scoped within core AI/ML conferences is becoming narrow, raising the question as to where the work fits best.

The IEEE Transactions on Software Engineering editorial board has recently discussed 
such challenges \cite{UchitelCPAABBBCDFHHLLMM24}.
While we do not have easy answers to these challenges, we conjecture that the complexity, scale, and pervasiveness of modern software systems---including AI-intensive ones---will make SE research more important than ever. At the same time, we also believe that SE research should not become a simple application of AI techniques in the software development domain. We believe that SE expertise, domain knowledge, and contributions are crucial to advance future SE activities that are becoming more influenced by AI in a truly effective, human-centric, sustainable, and trustworthy manner, and this will allow the SE community to keep its key role and identity.




\balance

\bibliographystyle{ACM-Reference-Format}
\bibliography{bibliography}


\end{document}